\documentstyle[preprint,aps]{revtex}
\input epsf.tex
\def\DESepsf(#1 width #2){\epsfxsize=#2 \epsfbox{#1}}

\newlength{\dinwidth}
\newlength{\dinmargin}
\begin{document}
\preprint{\vbox{\hbox{hep-ph/0108054}}} \draft
\title {Hadronic $B$ Decays to Charmless $VT$ Final States}
\author{C. S. Kim\footnote{cskim@mail.yonsei.ac.kr,~~~
http://phya.yonsei.ac.kr/\~{}cskim/}, B. H.
Lim\footnote{bhlim@yonsei.ac.kr}, and Sechul
Oh\footnote{scoh@phya.yonsei.ac.kr}}
\address{Department of Physics and IPAP, Yonsei University, Seoul,
120-749, Korea}
 \maketitle
\begin{abstract}
\noindent Charmless hadronic decays of $B$ mesons to a vector
meson ($V$) and a tensor meson ($T$) are analyzed in the
frameworks of both flavor SU(3) symmetry and generalized
factorization.  We also make comments on $B$ decays to two tensor
mesons in the final states. Certain ways to test validity of the
generalized factorization are proposed, using $B \to VT$ decays.
We calculate the branching ratios and CP asymmetries using the
\emph{full} effective Hamiltonian including all the \emph{penguin}
operators and the form factors obtained in the non-relativistic
quark model of Isgur, Scora, Grinstein and Wise.
\end {abstract}

\newpage
\section{Introduction}

In the next few years $B$ factories operating at KEK and SLAC will
provide plenty of new experimental data on $B$ decays. It is
expected that improved new bound will be put on the branching
ratios for various decay modes and many decay modes with small
branching ratios will be observed for the first time.  Thus more
information on rare decays of $B$ mesons will be available soon.
Experimentally several tensor mesons have been observed \cite{4A},
such as the isovector $a_2$(1320), the isoscalars $f_2$(1270),
$f_2^{\prime}$(1525), $f_2$(2010), $f_2$(2300), $f_2$(2340),
$\chi_{c2}(1P)$, $\chi_{b2}(1P)$ and $\chi_{c2}(2P)$, the
isospinors $K_2^*$(1430) and $D_2^*$(2460).  Experimental data on
the branching ratios for $B$ decays involving a vector ($V$) and a
tensor meson ($T$) in the final state provide only upper bounds,
as follows \cite{4A}:
\begin{eqnarray}
{\cal B} (B^{+} \rightarrow \rho^+ D_2^*(2460)^{0}) &<& 4.7 \times
10^{-3},   \nonumber
\\ {\cal B} (B^{0} \rightarrow \rho^+ D_2^*(2460)^{-}) &<&
4.9 \times 10^{-3},   \nonumber
\\ {\cal B} (B^{+} \rightarrow \rho^0 K_2^*(1430)^{+}) &<& 1.5
\times 10^{-3},   \nonumber
\\ {\cal B} (B^{0} \rightarrow \rho^0 K_2^*(1430)^{0}) &<&
1.1 \times 10^{-3},   \nonumber
\\ {\cal B} (B^{+} \rightarrow \phi K_2^*(1430)^{+}) &<& 3.4
 \times 10^{-3},   \nonumber
\\ {\cal B} (B^{0} \rightarrow \phi K_2^*(1430)^{0}) &<&
1.4 \times 10^{-3},   \nonumber
\\ {\cal B} (B^+ \rightarrow \rho^0 a_2(1320)^+) &<& 7.2 \times
10^{-4}.
\end{eqnarray}
In particular, the process $B \to K_2^* \gamma$ has been observed
for the first time by the CLEO Collaboration with a branching
ratio of $(1.66^{+0.59}_{-0.53} \pm 0.13) \times 10^{-5}$
\cite{cleo}.

There have been a few works \cite{1,2,3} studying two-body
hadronic $B$ decays involving a tensor meson $T$ ($J^P = 2^+$) in
the final state using the non-relativistic quark model of Isgur,
Scora, Grinstein and Wise (ISGW) \cite{4} with the factorization
ansatz.  However, those works considered only the tree diagram
contribution even in charmless $B$ decays to $PT$ ($P$ denotes a
pseudoscalar meson) and $VT$, such as $B \rightarrow
\eta^{(\prime)} a_2$ and $B \rightarrow \phi f_2^{(\prime)}$.  In
most cases of the charmless $\Delta S =0$ processes, the dominant
contribution arises from the tree diagram and the contributions
from the penguin diagrams are very small. But in some cases such
as $B \rightarrow \eta^{(\prime)} a_2$ and $\eta^{(\prime)}
f_2^{(\prime)}$, the penguin diagrams could provide sizable
contributions. Furthermore, in the charmless $|\Delta S| =1$ decay
processes, the penguin diagram contribution is enhanced by the CKM
matrix elements $V^*_{tb} V_{ts}$ and becomes dominant.

In a recent work \cite{btopt}, we have studied $B$ decays to a
pseudoscalar meson and a tensor meson. In this work, the previous
analysis is extended to charmless hadronic decays of $B$ mesons to
a vector meson and a tensor meson in the frameworks of \emph{both}
flavor SU(3) symmetry and the generalized factorization. We also
comment on $B$ decays to \emph{two} tensor mesons in the final
states. Purely based on the flavor SU(3) symmetry, we first
present a model-independent analysis in $B \rightarrow VT$ decays.
Then we use the \emph{full} effective Hamiltonian including all
the penguin operators and the ISGW quark model to calculate the
branching ratios for $B \rightarrow VT$ decays. Since we include
both the tree and the penguin diagram contributions to decay
processes, we are able to calculate the branching ratios for all
the charmless $|\Delta S| =1$ decays and the relevant CP
asymmetries.  In order to bridge the flavor SU(3) approach and the
factorization approach, we present a set of relations between a
flavor SU(3) amplitude and a corresponding amplitude in the
factorization in $B \rightarrow VT$ decays. Certain ways to test
validity of the generalized factorization are proposed by
emphasizing interplay between both approaches.

This work is organized as follows.  In Sec. II we discuss the
notations for SU(3) decomposition and the full effective
Hamiltonian for $B$ decays. We also make some comments on $B \to
TT$ decays in Sec. II~. In Sec. III we present a model-
independent analysis of $B \rightarrow VT$ decays based on SU(3)
symmetry.  In Sec. IV the two-body decays $B \rightarrow VT$ are
analyzed in the framework of generalized factorization.  The
branching ratios and CP asymmetries are calculated using the form
factors obtained in the ISGW quark model.  Finally, in Sec. V we
conclude our analysis.

\section{Framework}

Since in $B \to VT$ decays there are three possible partial waves
with $l =1,2,3$ in the final state, $B \to VT$ processes are more
complicated than $B \to PT$ processes. For the SU(3) analysis of
$B \to VT$ decays, these partial waves in the final state need to
be separated out.  We will assume that this can be done by certain
methods such as one using angular distributions in $B \to VV$
decays \cite{dighe}. In the flavor SU(3) approach, the decay
amplitudes of two-body $B$ decays are decomposed into linear
combinations of the SU(3) amplitudes, which are reduced matrix
elements defined in Ref. \cite{5}.  In SU(3) decomposition of
decay amplitudes of the $B \rightarrow VT$ processes, we choose
the notations given in Refs. \cite{5,7,8} as follows: We represent
the decay amplitudes in terms of the basis of quark diagram
contributions, $T$ (tree), $C$ (color-suppressed tree), $P$
(QCD-penguin), $S$ (additional penguin effect involving
SU(3)-singlet mesons), $E$ (exchange), $A$ (annihilation), and
{\it PA} (penguin annihilation). The amplitudes $E$, $A$ and {\it
PA} may be neglected to a good approximation because of a
suppression factor of $f_B / m_B \approx 5 \%$.  For later
convenience we also denote the electroweak (EW) penguin effects
explicitly as $P_{EW}$ (color-favored EW penguin) and $P_{EW}^{C}$
(color-suppressed EW penguin), even though in terms of quark
diagrams the inclusion of these EW penguin effects only leads to
the following replacement without introducing new SU(3)
amplitudes; $T \rightarrow T + P_{EW}^{C}$, $C \rightarrow C +
P_{EW}$, $P \rightarrow P -{1 \over 3} P_{EW}^{C}$, $S \rightarrow
S -{1 \over 3} P_{EW}$.  We use the following phase convention for
the vector and the tensor mesons:
\begin{eqnarray}
\rho^+ (a^+_2) &=& u \bar d~, \;\;\; \rho^0 (a^0_2) = -{1 \over
\sqrt{2}} (u \bar u -d \bar d)~, \;\;\; \rho^- (a^-_2) = -\bar u
d~, \nonumber
\\ K^{*+} (K^{*+}_2) &=& u \bar s~, \;\;\; K^{*0} (K^{*0}_2) = d \bar s~,
\;\;\; \bar K^{*0} (\bar K^{*0}_2) = \bar d s~, \;\;\; K^{*-}
(K^{*-}_2) = - \bar u s~,  \nonumber
\\ \omega &=& {1 \over \sqrt{2}} (u \bar u + d \bar d)~,
\;\; \phi = s \bar s~, \nonumber
\\ f_2 &=& {1 \over \sqrt{2}} (u \bar u +d \bar d) \cos \phi_{_T} +(s
\bar s) \sin \phi_{_T} ~, \;\; f^{\prime}_2 = {1 \over \sqrt{2}}
(u \bar u +d \bar d) \sin \phi_{_T} -(s \bar s) \cos \phi_{_T} ~,
\end{eqnarray}
where the mixing angle $\phi_{_T}$ is given by $\phi_{_T} =
\arctan(1 / \sqrt{2}) - 28^0 \approx 7^0$ \cite{1,8A}.

In the factorization scheme, we first consider the effective weak
Hamiltonian. We then use the generalized factorization
approximation to derive hadronic matrix elements by saturating the
vacuum state in all possible ways.  The method includes color
octet non-factorizable contribution by treating $\xi\equiv 1/N_c$
($N_c$ denotes the effective number of color) as an adjustable
parameter.  The generalized factorization approximation has been
quite successfully used in two-body $D$ decays as well as $B
\rightarrow D$ decays\cite{9}. The effective weak Hamiltonian for
hadronic $\Delta B=1$ decays can be written as
\begin{eqnarray}
 H_{eff} &=& {4 G_{F} \over \sqrt{2}} \left[ V_{ub}V^{*}_{uq} (c_1
O^{u}_1 +c_2 O^{u}_2)
   + V_{cb}V^{*}_{cq} (c_1 O^{c}_1 +c_2 O^{c}_2)
   - V_{tb}V^{*}_{tq} \sum_{i=3}^{12} c_{i} O_{i} \right] \nonumber \\
  &+& {\rm H.C.} ~,
\end{eqnarray}
where $O_{i}$'s are defined as
\begin{eqnarray}
O^f_1 &=& (\bar q \gamma_{\mu} L f) (\bar f \gamma^{\mu} L b) ~,
\;\; O^f_2 = (\bar q_{\alpha} \gamma_{\mu} L f_{\beta}) (\bar
f_{\beta} \gamma^{\mu} L b_{\alpha})~,    \nonumber \\ O_{3(5)} &=&
(\bar q \gamma_{\mu} L b) (\Sigma \bar q^{\prime} \gamma^{\mu}
L(R) q^{\prime})~,  \;\; O_{4(6)} = (\bar q_{\alpha} \gamma_{\mu} L
b_{\beta}) (\Sigma \bar q^{\prime}_{\beta} \gamma^{\mu} L(R)
q^{\prime}_{\alpha})~,  \nonumber \\ O_{7(9)} &=& {3 \over 2} (\bar
q \gamma_{\mu} L b) (\Sigma e_{q^{\prime}} \bar q^{\prime}
\gamma^{\mu} R(L) q^{\prime})~,  \;\; O_{8(10)} ={3 \over 2} (\bar
q_{\alpha} \gamma_{\mu} L b_{\beta}) (\Sigma e_{q^{\prime}} \bar
q^{\prime}_{\beta} \gamma^{\mu} R(L) q^{\prime}_{\alpha})~,
\nonumber \\ O_{11} &=& {g_s \over 32 \pi^2} m_b (\bar q
\sigma^{\mu \nu} R T^a b) G^a_{\mu \nu}~,  \;\; O_{12} = {e \over
32 \pi^2} m_b (\bar q \sigma^{\mu \nu} R b) F_{\mu \nu}~.
\end{eqnarray}
Here $c_i$'s are the Wilson coefficients (WC's) evaluated at the
renormalization scale $\mu$.  And $L(R) = (1 \mp \gamma_5)/2$, $f$
can be $u$ or $c$ quark, $q$ can be $d$ or $s$ quark, and
$q^{\prime}$ is summed over $u$, $d$, $s$, and $c$ quarks.
$\alpha$ and $\beta$ are the SU(3) color indices, and $T^a$ ($a=
1,...,8$) are the SU(3) generator with the normalization $Tr (T^a
T^b) =\delta^{ab} /2$.  $g_s$ and $e$ are the strong and electric
couplings, respectively.  $G^a_{\mu \nu}$ and $F_{\mu \nu}$ denote
the gluonic and photonic field strength tensors, respectively.
$O_1$ and $O_2$ are the tree-level and QCD-corrected operators.
$O_{3-6}$ are the gluon-induced strong penguin operators.
$O_{7-10}$ are the EW penguin  operators due to $\gamma$ and $Z$
exchange, and box diagrams at loop level.  We shall take into
account the chromomagnetic operator $O_{11}$ but neglect the
extremely small contribution from $O_{12}$. The dipole
contribution is in general quite small, and is of the order of
$10\%$ for penguin dominated modes. For all the other modes it can
be neglected \cite{10}.

We use the ISGW quark model to analyze two-body charmless decay
processes $B \rightarrow VT$  in the framework of generalized
factorization.  We describe the parameterizations of the hadronic
matrix elements in $B \rightarrow VT$ decays \cite{4}:
\begin{eqnarray}
\langle 0 | V^{\mu} | V \rangle &=& f_{_V} m_{_V} \epsilon^{\mu} ~,  \\
\langle T | j^{\mu} | B \rangle &=& i h(m_P^2) \epsilon^{\mu \nu
\rho \sigma} \epsilon^*_{\nu \alpha} p_B^{\alpha} (p_B
+p_T)_{\rho} (p_B -p_T)_{\sigma} + k(m_P^2) \epsilon^{* \mu \nu}
(p_B)_{\nu}  \nonumber \\
&\mbox{}&  + \epsilon^*_{\alpha \beta} p_B^{\alpha} p_B^{\beta} [
b_+(m_P^2) (p_B +p_T)^{\mu} +b_-(m_P^2) (p_B -p_T)^{\mu} ]~,
\label{formfactor}
\end{eqnarray}
where $j^{\mu} = V^{\mu} -A^{\mu}$.  $V^{\mu}$ and $A^{\mu}$
denote a vector and an axial-vector current, respectively.  $f_P$
denotes the decay constant of the relevant pseudoscalar meson.
$h(m_P^2)$, $k(m_P^2)$, $b_+(m_P^2)$, and $b_-(m_P^2)$ express the
form factors for the $B \rightarrow T$ transition,
$F^{B \rightarrow T}(m_P^2)$, which have been
calculated at $q^2 =m_P^2$ $(q^{\mu} \equiv p_B^{\mu} -p_T^{\mu})$
in the ISGW quark model \cite{4}. $p_B$ and $p_T$ denote the
momentum of the $B$ meson and the tensor meson, respectively.

The polarization tensor $\epsilon^{\mu \nu}$ of the tensor meson
$T$ satisfies the following properties \cite{epsilon}:
\begin{eqnarray}
&\mbox{}& \epsilon^{\mu \nu} (p_{_T}, \lambda) = \epsilon^{\nu
\mu} (p_{_T}, \lambda),
\\ &\mbox{}& p_\mu \epsilon^{\mu \nu} (p_{_T}, \lambda) = p_\nu \epsilon^{\mu \nu} (p_{_T},
\lambda) = 0,
\\ &\mbox{}& \epsilon^\mu_{~\mu} (p_{_T}, \lambda) = 0~,
\end{eqnarray}
where $\lambda$ is the helicity index of the tensor meson.  We
note that due to the above properties of the polarization tensor,
the matrix element $\langle 0 | j^{\mu} | T \rangle$ vanishes:
\begin{equation}
\langle 0 | j^{\mu} | T \rangle = p_\nu \epsilon^{\mu \nu}
(p_{_T}, \lambda) + p_{_T}^\mu \epsilon^\nu_{~\nu} (p_{_T},
\lambda) =0~. \label{fT}
\end{equation}
Thus, in the generalized factorization scheme, just as in the case
of $B \to PT$ decays, the decay amplitudes for $B \rightarrow VT$
can be considerably simplified, compared to those for other
two-body charmless decays of $B$ mesons such as $B \rightarrow
PP$, $PV$, and $VV$: Any decay amplitude for $B \rightarrow VT$ is
simply proportional to the decay constant $f_V$ and a certain
linear combination of the form factors $F^{B \rightarrow T}$, {\it
i.e.}, there is no such amplitude proportional to $f_T \times
({\rm form~ factor~ for~} B \rightarrow V)$.

We would like to make comments on decays of $B$ mesons to two
tensor mesons in the final state.  Since $\langle 0 | j_{\mu} | T
\rangle = 0$, in the factorization scheme the decay amplitude for
$B \to TT$ decays always vanishes:
\begin{eqnarray}
\langle TT | H_{eff} | B \rangle \sim  \langle T | j^{\mu} | B
\rangle ~  \langle 0 | j_{\mu} | T \rangle = 0
\end{eqnarray}
Non-zero of a rate for any $B \to TT$ decay would arise from
non-factorizable effects or final state interactions.  Therefore,
search for any $B \to TT$ modes in future experiment can provide a
critical test of the factorization ansatz.

\section{Flavor SU(3) analysis of $B \rightarrow VT$ decays}

We list the $B \rightarrow VT$ decay modes in terms of the SU(3)
amplitudes.  The coefficients of the SU(3) amplitudes in $B
\rightarrow VT$ are listed in Tables I and II for
strangeness-conserving ($\Delta S =0$) and strangeness-changing
($|\Delta S| =1$) processes, respectively. In the tables, the
unprimed and the primed letters denote $\Delta S =0$ and $|\Delta
S| =1$ processes, respectively. The subscript, $V$ in $T_V,~C_V,$
... or $T$ in $T_T,C_T,$ ..., on each SU(3) amplitude is used to
describe such a case that the meson, which includes the spectator
quark in the corresponding quark diagram, is the vector $V$ or the
tensor $T$.  Note that the coefficients of the SU(3) amplitudes
with the subscript $V$, which would be proportional to $f_T \times
F^{B \rightarrow V}$, are expressed in square brackets. As
explained in Sec. II, the contributions of the SU(3) amplitudes
with the subscript $V$ vanish in the framework of factorization,
because those contributions contain the matrix element $\langle T|
J^{\rm{weak}}_{\mu} |0 \rangle$ which is zero, see Eq. (\ref{fT}).
Thus, it will be interesting to compare the results obtained in
the SU(3) analysis with those obtained in the factorization
scheme, as we shall see.  We will present some ways to test
validity of both schemes in future experiment.

Among the $\Delta S =0$ amplitudes, the tree diagram contribution
is expected to be largest so that from Table I the decays $B^+
\rightarrow \rho^+ a_2^0$, $\rho^+ f_2$, and $B^0 \rightarrow
\rho^+ a_2^-$ are expected to have the largest rates.  Here we
have noticed that in $B^+ \rightarrow \rho^+ f_2^{(\prime)}$
decays, $\cos \phi_{_T} =0.99$ and $\sin \phi_{_T} =0.13$, since
the mixing angle $\phi_{_T} \approx 7^0$. The amplitudes for the
processes $B \to \phi f_2^{(\prime)}$, $\phi a_2$, and $K^* K^*_2$
have only penguin diagram contributions, and so they are expected
to be small.  In principle, the penguin contribution (combined
with the smaller color-suppressed EW penguin) $p_T \equiv P_T -{1
\over 3} P_{EW,T}$ can be measured in $B^{+(0)} \rightarrow \bar
K^{*0} K_2^{*+(0)}$.  The tree contribution (combined with much
smaller color-suppressed EW penguin) $t_T \equiv T_T +P^C_{EW,T}$
are measured by the combination $A(B^{+(0)} \rightarrow \bar
K^{*0} K_2^{*+(0)}) -A(B^0 \rightarrow \rho^+ a_2^-)$.  The
amplitudes for $B^0 \rightarrow \rho^0 f_2^{\prime}$ and $\omega
f_2^{\prime}$ have the color-suppressed tree contributions, $C_T
(C_V)$, but are suppressed by $\sin \phi_{_T}$ so that they are
expected to be small. We shall see that these expectations based
on the SU(3) approach are consistent with those calculated in the
factorization approximation. However, there exist some cases in
which the predictions based on both approaches are inconsistent.
Note that in Table I the amplitudes for $B^0 \rightarrow \rho^-
a_2^+$ and $B^{+ (0)} \rightarrow K^{*+(0)} \bar K^{*0}_2$ can be
decomposed into linear combinations of the SU(3) amplitudes as
follows:
\begin{eqnarray}
A(B^0 \rightarrow \rho^- a_2^+) &=& - T_V -P_V -(2/3) P^C_{EW,V}~,
\label{b0pia2}  \\
A(B^+ \rightarrow K^{*+} \bar K^{*0}_2) &=& A(B^0 \rightarrow
K^{*0} \bar K^{*0}_2) = P_V -(1/3) P^C_{EW,V}~.
\end{eqnarray}
As previously explained, in factorization the rates for these
processes vanish because all the SU(3) amplitudes are with the
subscript $V$. Non-zero of decay rates for these processes would
arise from non-factorizable effects or final state interactions.
Thus, in principle one can test validity of the factorization
ansatz by measuring the rates for these decays in future
experiment.  Furthermore, the non-factorizable penguin
contribution, if exists, (combined with the smaller
color-suppressed EW penguin) $p_V \equiv P_V -{1 \over 3}
P_{EW,V}$ can be measured in $B^{+(0)} \rightarrow \bar K^{*+(0)}
\bar K_2^{*+(0)}$.  Also, supposing that $P_V$ is very small
compared to $T_V$ as usual, one can determine the magnitude of
$T_V$ by measuring the rate for $B^0 \rightarrow \rho^- a^+_2$.

In the $|\Delta S|=1$ decays, the (strong) penguin contribution
$P^{\prime}$ is expected to dominate because of enhancement by the
ratio of the CKM elements $|V^*_{tb} V_{ts}| / |V^*_{ub} V_{us}|
\approx 50$.  We note that the amplitudes for $B^+ \rightarrow
K^{*0} a_2^+$ and $B^+ \rightarrow \rho^+ K_2^{*0}$ have only
penguin contributions, respectively, as follows:
\begin{eqnarray}
A (B^+ \rightarrow K^{*0} a_2^+) &=& P_T^{\prime} -{1 \over 3}
P_{EW,T}^{C \prime}~, \\
A (B^+ \rightarrow \rho^+ K_2^{*0}) &=& P_V^{\prime} -{1 \over 3}
P_{EW,V}^{C \prime}~. \label{piK2}
\end{eqnarray}
Thus the penguin contribution (combined with the smaller
color-suppressed EW penguin) $p_T^{\prime} \equiv P_T^{\prime} -{1
\over 3} P_{EW,T}^{C \prime}$ is measured in $B^+ \rightarrow
K^{*0} a_2^+$.  Similarly, $p_V^{\prime} \equiv P_V^{\prime} -{1
\over 3} P_{EW,V}^{C \prime}$ is determined in $B^+ \rightarrow
\rho^+ K_2^{*0}$. (In fact, $p_V^{\prime} =0$ in factorization.)
By comparing the branching ratios for these two modes measured in
experiment, one can determine which contribution (i.e.,
$p_T^{\prime}$ or $p_V^{\prime}$) is larger. The (additional
penguin) SU(3) singlet amplitude $S^{\prime}$ is expected to be
very small because of the Okubo-Zweig-Iizuka (OZI) suppression. As
in $\Delta S=0$ decays, there are certain processes whose
amplitudes can be expressed by the SU(3) amplitudes, but are
expected to vanish in factorization:  For instance, $A (B^+
\rightarrow \rho^+ K_2^{*0})$ is given by Eq. (\ref{piK2}) and $A
(B^0 \rightarrow \rho^- K_2^{*+}) = -(T_V^{\prime} +P_V^{\prime}
+{2 \over 3} P_{EW,V}^{C \prime})$.  Thus, in principle
measurement of the rates for these decays can be used to test the
factorization ansatz.  We also note that the decay amplitudes for
modes $B^+ \rightarrow \rho^0 K_2^{*+}$ and $B^0 \rightarrow
\rho^0 K_2^{*0}$ can be respectively written as
\begin{eqnarray}
A (B^+ \rightarrow \rho^0 K_2^{*+}) &=& -{1 \over \sqrt{2}}
(T_V^{\prime} + C_T^{\prime} +P_V^{\prime} +P_{EW,T}^{\prime} +{2
\over 3} P_{EW,V}^{C \prime})~, \\
A (B^0 \rightarrow \rho^0 K_2^{*0}) &=& -{1 \over \sqrt{2}} (
C_T^{\prime} -P_V^{\prime} +P_{EW,T}^{\prime} +{1 \over 3}
P_{EW,V}^{C \prime})~.
\end{eqnarray}
Since in factorization only the amplitudes having the subscript
$T$ does not vanish, we shall see that ${\cal B} (B^+ \rightarrow
\rho^0 K_2^{*+}) = {\cal B} (B^0 \rightarrow \rho^0 K_2^{*0})$ in
the factorization scheme, where ${\cal B}$ denotes the branching
ratio.  Thus, if $T_V^{\prime}$ or $P_V^{\prime}$ is (not zero
and) not very suppressed compared to $C_T^{\prime}$, then there
would be a sizable discrepancy in the relation ${\cal B} (B^+
\rightarrow \rho^0 K_2^{*+}) = {\cal B} (B^0 \rightarrow \rho^0
K_2^{*0})$, and in principle it can be tested in experiment.

{}From Tables I and II, we find some useful relations among the
decay amplitudes.  The equivalence relations are: for the $\Delta
S = 0$ modes,
\begin{eqnarray}
{1 \over \sqrt{2}}A(B^+ \rightarrow \phi a^{+}_2) &=& A(B^0
\rightarrow \phi a^{0}_2 ) ~,  \nonumber
\\ = {1 \over c} A(B^0 \rightarrow \phi f_2) &=& {1 \over s}
A(B^0 \rightarrow \phi f^{\prime}_2 ) ~, \nonumber
\\ A(B^+ \rightarrow K^{*+} \bar K^{*0}_2) &=& A(B^0 \rightarrow
K^{*0} \bar K^{*0}_2)~, \nonumber
\\ A(B^+ \rightarrow \bar K^{*0} K^{*+}_2) &=& A(B^0 \rightarrow \bar
K^{*0} K^{*0}_2)~,
\end{eqnarray}
and for the $|\Delta S| = 1$ modes,
\begin{eqnarray}
A(B^+ \rightarrow \phi K^{*+}_2) &=& A(B^0 \rightarrow \phi
K^{*0}_2 )~.
\end{eqnarray}
The quadrangle relations are: for the $\Delta S = 0$ processes,
\begin{eqnarray}
{1 \over c} A(B^+ \rightarrow \rho^+  f_2)- {1 \over s}A(B^+
\rightarrow \rho^+ f^{\prime}_2)  &=& \sqrt{2} \left[{1 \over c}
A(B^0 \rightarrow \rho^0  f_2)- {1 \over s}A(B^0 \rightarrow
\rho^0 f^{\prime}_2) \right]  \nonumber
\\ &=& \sqrt{2} \left[{1 \over c} A(B^0 \rightarrow \omega f_2)-{1 \over s}
A(B^0 \rightarrow \omega f^{\prime}_2) \right] ~, \label{deltas0}
\end{eqnarray}
and for the $|\Delta S| = 1$ processes,
\begin{eqnarray}
A(B^+ \rightarrow  K^{* 0} a^{+}_2 ) + \sqrt{2} A(B^+ \rightarrow
K^{* +} a^{0}_2) &=& \sqrt{2} A(B^0 \rightarrow K^{* 0} a^{0}_2 )
+ A(B^0 \rightarrow K^{* +} a^{-}_2) ~, \nonumber
\\ {1 \over c}
A(B^+ \rightarrow  K^{* +} f_2)- {1 \over s}A(B^+ \rightarrow K^{*
+} f^{\prime}_2) &=& {1 \over c} A(B^0 \rightarrow K^{* 0} f_2)-{1
\over s} A(B^0 \rightarrow K^{* 0} f^{\prime}_2) ~, \nonumber
\\ A(B^+ \rightarrow  \rho^+ K^{* 0} ) + \sqrt{2} A(B^+ \rightarrow
\rho^0 K^{* +}) &=&  A(B^0 \rightarrow \rho^- K^{* +}) + \sqrt{2}
A(B^0 \rightarrow \rho^0 K^{* 0}_2) ~,  \label{deltas1}
\end{eqnarray}
where $c \equiv \cos \phi_{_T}$ and $s \equiv \sin \phi_{_T}$.
Note that the above relations are derived, purely based on flavor
SU(3) symmetry.  In the factorization scheme, (neglecting the
SU(3) amplitudes with the subscript $V$) we would have in addition
the approximate relations as follows.\footnote{Considering SU(3)
breaking effects, we use the symbol $\approx$ in the following
relations instead of the equivalence symbol $=$.} The following
factorization relation would hold:
\begin{eqnarray}
\sqrt{2} A(B^+ \rightarrow \rho^+ a_2^0) &\approx& A(B^0
\rightarrow \rho^+ a_2^-)~.  \label{facto1}
\end{eqnarray}
The quadrangle relations given in Eqs. (\ref{deltas0},
\ref{deltas1}) would be divided into the following factorization
relations: for the $\Delta S = 0$ processes,
\begin{eqnarray}
{1 \over c} A(B^+ \rightarrow \rho^+  f_2) &\approx& {1 \over
s}A(B^+ \rightarrow \rho^+ f^{\prime}_2)~,  \nonumber
\\ {1 \over c} A(B^0 \rightarrow \rho^0  f_2) &\approx& {1 \over s}A(B^0
\rightarrow \rho^0 f^{\prime}_2)~, \nonumber
\\  {1 \over c} A(B^0 \rightarrow \omega f_2) &\approx& {1 \over s}
A(B^0 \rightarrow \omega f^{\prime}_2) ~,  \label{facto2}
\end{eqnarray}
and for the $|\Delta S| = 1$ processes,
\begin{eqnarray}
\sqrt{2} A(B^+ \rightarrow K^{*+} a_2^0) &\approx& A(B^0
\rightarrow K^{*+} a_2^-)~,  \nonumber
\\ A(B^+ \rightarrow K^{*0} a^+_2) &\approx& \sqrt{2} A(B^0 \rightarrow
K^{*0} a^0_2)~,  \nonumber
\\ {1 \over c} A(B^+ \rightarrow K^{*+} f_2) &\approx& {1 \over s} A(B^+
\rightarrow K^{*+} f^{\prime}_2)~,  \nonumber
\\ {1 \over c} A(B^0 \rightarrow K^{*0} f_2) &\approx& {1 \over s}
A(B^0 \rightarrow K^{*0} f^{\prime}_2)~,  \nonumber
\\ A(B^+ \rightarrow \rho^0 K^{*+}_2) &\approx&
A(B^0 \rightarrow \rho^0 K^{*0}_2)~,  \nonumber
\\ A(B^+ \rightarrow \omega K^{*+}_2) &\approx& A(B^0
\rightarrow \omega K^{*0}_2)~.  \label{facto3}
\end{eqnarray}
Therefore, in principle the above relations given in Eqs.
(\ref{facto1}, \ref{facto2}, \ref{facto3}) provide an interesting
way to test the factorization scheme by measuring and comparing
magnitudes of the decay amplitudes involved in the relations.  In
consideration of SU(3) breaking effects, the relation in Eq.
(\ref{facto1}) is best to use, because in fact the relation arises
from isospin symmetry assuming $C_V = P_V = P_{EW,V} =P^C_{EW,V}
=0$.  (However, if $C_V$ is negligibly small (though not zero)
compared to $T_T$, Eq. (\ref{facto1}) will approximately hold.)

\section{Analysis of $B \rightarrow VT$ in the Isgur-Scora-Grinstein-Wise model}

We present a set of relations between a flavor SU(3) amplitude
involved in $B \rightarrow VT$ decays and a corresponding
amplitude in the generalized factorization, which bridge both
approaches in $B \rightarrow VT$ decays as follows \cite{12}.
(Note that all the SU(3) amplitudes with the subscript $P$, such
as $T^{(\prime)}_P$ etc., vanish because those are proportional to
the matrix element $\langle T | j^{\mu} | 0 \rangle$.)
\begin{eqnarray}
T^{(\prime)}_T &=& i {G_F \over \sqrt{2}} V^*_{ub} V_{ud (s)} (
m_{V}  f_{V} \epsilon^{* \alpha \beta} F_{\alpha \beta}^{B
\rightarrow {T}} (m^2_{V}))
a_1~, \nonumber \\
C^{(\prime)}_T &=& i {G_F \over \sqrt{2}} V^*_{ub} V_{ud (s)} (
m_{V}  f_{V} \epsilon^{* \alpha \beta} F_{\alpha \beta}^{B
\rightarrow {T}} (m^2_{V}))
a_2~,  \nonumber \\
S^{(\prime)}_T &=& -i {G_F \over \sqrt{2}} V^*_{tb} V_{td (s)} (
m_{V}  f_{V} \epsilon^{* \alpha \beta} F_{\alpha \beta}^{B
\rightarrow {T}} (m^2_{V}))
(a_3 +a_5)~,  \nonumber \\
P^{(\prime)}_T &=& -i {G_F \over \sqrt{2}} V^*_{tb} V_{td (s)} (
m_{V}  f_{V} \epsilon^{* \alpha \beta} F_{\alpha \beta}^{B
\rightarrow {T}} (m^2_{V}))
a_4 ~,  \nonumber \\
P_{EW,T}^{(\prime)} &=& -i {G_F \over \sqrt{2}} V^*_{tb} V_{td
(s)} ( m_{V}  f_{V} \epsilon^{* \alpha \beta} F_{\alpha \beta}^{B
\rightarrow {T}} (m^2_{V}))
{3 \over 2} (a_7 +a_9)~, \nonumber \\
P_{EW,T}^{C (\prime)} &=& -i {G_F \over \sqrt{2}} V^*_{tb} V_{td
(s)} ( m_{V}  f_{V} \epsilon^{* \alpha \beta} F_{\alpha \beta}^{B
\rightarrow {T}} (m^2_{V})) {3 \over 2} a_{10}~, \label{su3facto}
\end{eqnarray}
where
\begin{eqnarray}
 F_{\alpha \beta}^{B \rightarrow T}(m_V^2) &=&
 {\epsilon_{\mu}}^* (p_{_B} +p_{_T})_{\rho} [ i h(m_V^2) \cdot \epsilon^{\mu \nu
 \rho \sigma } {g_{\alpha \nu}} (p_{_V})_{\beta} (p_{_V})_{\sigma}
 +k(m_V^2) \cdot {\delta^{\mu}}_{\alpha}  {\delta^{\rho}}_{\beta}  \nonumber
\\ &\mbox{}& + b_+(m_V^2) \cdot
 {(p_{_V})}_{\alpha} {(p_{_V})}_{\beta} g^{\mu \rho}]. \label{FBT}
\end{eqnarray}
Here the effective coefficients $a_i$ are defined as $a_i =
c^{eff}_i + \xi c^{eff}_{i+1}$ ($i =$ odd) and $a_i = c^{eff}_i +
\xi c^{eff}_{i-1}$ ($i =$ even) with the effective WC's
$c^{eff}_i$ at the scale $m_b$ \cite{10}, and by treating
$\xi\equiv 1/N_c$ ($N_c$ denotes the effective number of color) as
an adjustable parameter.

With Tables I, II and the above relations (\ref{su3facto}), one
can easily write down in the factorization scheme the amplitude of
any $B \rightarrow VT$ mode shown in the tables.  For example,
from Table I and the relations (\ref{su3facto}), the amplitude of
the process $B^+ \rightarrow \rho^+ a_2^0$ can be written
as\footnote{In the factorization scheme, we use the usual phase
convention for the pseudoscalar and the tensor mesons as follows:
$\rho^0 (a^0_2) = {1 \over \sqrt{2}} (u \bar u -d \bar d)$,
$\rho^- (a^-_2) = \bar u d$,  $K^{*-} (K^{*-}_2) = \bar u s$. }
\begin{eqnarray}
A (B^+ \rightarrow \rho^+ a^0_2) &=& -{1 \over \sqrt{2}} \left(T_T
+ C_V +P_T -P_V +P_{EW,V} +{2 \over 3} P^C_{EW,T} +{1 \over 3}
P^C_{EW,V} \right) \nonumber \\
&=& {G_F \over \sqrt{2}} ( m_{\rho^+}  f_{\rho^+} \epsilon^{*
\alpha \beta} F_{\alpha \beta}^{B \rightarrow {a_2^{0}}}
(m^2_{\rho^+})) [ V^*_{ub} V_{ud} a_1 - V^*_{tb} V_{td} (a_4 +
a_{10}) ] ~.
\end{eqnarray}
Here we have used the fact that $C_V$, $P_V$, $P_{EW,V}$, and
$P^C_{EW,V}$ with the subscript $V$ all vanish in factorization.
Expressions for all the amplitudes of $B \rightarrow VT$ decays
are given in Appendix as calculated in the factorization scheme.

The unpolarized decay rate for $B \rightarrow VT$ is given by
\begin{eqnarray}
\Gamma (B \rightarrow VT) &=& {G_F^2 \over 48 \pi m_{_T}^4} m_{_V}
f^2_{_V} |\{ V^*_{ub} V_{ud(s)} \cdot (a_1~ {\rm or}~ a_2) -
V^*_{tb} V_{td(s)} \cdot (a_{i}{\rm 's}) \}|^2  \nonumber \\
&\mbox{}& \cdot [{\cal X} |\vec p_{_V}|^7 +{\cal Y} |\vec
p_{_V}|^5 +{\cal Z} |\vec p_{_V}|^3 ]~,
\end{eqnarray}
where $|\vec{p}_{_V}|$ is the magnitude of three-momentum of the
final state particle $V$ or $T$ ($|\vec{p}_{_V}|=|\vec{p}_{_T}|$)
in the rest frame of the $B$ meson.  The effective coefficients
$a_i$ are defined as in Eq. (\ref{su3facto}).  The factors ${\cal
X}$, ${\cal Y}$, and ${\cal Z}$, respectively, are given by
\begin{eqnarray}
{\cal X} &=& 8 m_{_B}^4 b_+^2~,  \nonumber \\
{\cal Y} &=& 2 m_{_B}^2 [ 6 m_{_V}^2 m_{_T}^2 h^2 +2 ( m_{_B}^2
-m_{_T}^2 -m_{_V}^2 ) k b_+ + k^2 ]~,  \nonumber \\
{\cal Z} &=& 5 m_{_T}^2 m_{_V}^2 k^2~.
\end{eqnarray}
Here we have summed over polarizations of the tensor meson $T$
using the following formula \cite{2}:
\begin{eqnarray}
\sum_{\lambda} \epsilon_{\alpha \beta}(p_{_T}, \lambda)
\epsilon^*_{\mu \nu}(p_{_T}, \lambda) = {1 \over 2}
(\theta_{\alpha \mu} \theta_{\beta \nu} +\theta_{\beta \mu}
\theta_{\alpha \nu}) -{1 \over 3} \theta_{\alpha \beta}
\theta_{\mu \nu}~,
\end{eqnarray}
where $\theta_{\alpha \beta} = - g_{\alpha \beta}
+(p_{_T})_{\alpha} (p_{_T})_{\beta}/ m_T^2$.

The CP asymmetry, ${\mathcal A}_{CP}$, is defined by
\begin{eqnarray}
{\mathcal A}_{CP} = {{\mathcal B}(B \rightarrow f)  -{\mathcal
B}(\bar B \rightarrow \bar f) \over {\mathcal B}(B \rightarrow f)
+{\mathcal B}(\bar B \rightarrow \bar f)}~,
\end{eqnarray}
where $B$ and $f$ denote $b$ quark and a generic final state,
respectively.

We calculate the branching ratios and CP asymmetries for $B
\rightarrow VT$ decay modes for various input parameter values.
The predictions are sensitive to several input parameters, such as
the form factors, the strange quark mass, the parameter $\xi
\equiv 1/ N_c$, the CKM matrix elements and in particular, the
weak phase $\gamma$. In a recent work \cite{10} on charmless $B$
decays to two light mesons such as $PP$ and $VP$, it has been
shown that the favored values of the input parameters are
$$
\xi \approx 0.45,~ m_s(m_b) \approx 85 ~{\rm MeV},~\gamma \approx
110^0,~ V_{cb} =0.040,~~~{\rm and}~~~ |V_{ub} /V_{cb}| =0.087
$$
in order to get the best fit to the recent
experimental data from the CLEO collaboration.  For our numerical
calculations, we use the following values of the decay constants
(in MeV units) \cite{9,13,14}:
$$
f_{\rho} =216,~~ f_{\omega} =216,~~ f_{\phi} =236,~~f_{K^*} =222.
$$
We use the values of the
form factors for the $B \rightarrow T$ transition calculated in
the ISGW model \cite{4}. The strange quark mass $m_s$ is in
considerable doubt: {\it i.e.}, QCD sum rules give $m_s(1\; {\rm
GeV})=(175\pm 25)$ MeV and lattice gauge theory gives $m_s(2\;
{\rm GeV})=(100 \pm 20 \pm 10)$ MeV in the quenched lattice
calculation \cite{15}. In this analysis we use two representative
values of $m_s = 100$ MeV and $m_s = 85$ MeV at $m_b$ scale.
Current best estimates for CKM matrix elements are $V_{cb} =0.0381
\pm 0.0021$ and $|V_{ub} /V_{cb}| =0.085 \pm 0.019$ \cite{16}.  We
use $V_{cb} =0.040$ and $|V_{ub} /V_{cb}| =0.087$.  It has been
known that there exists the discrepancy in values of $\gamma$
extracted from CKM-fitting at $\rho-\eta$ plane \cite{17} and from
the $\chi^2$ analysis of non-leptonic decays of $B$ mesons
\cite{18,19}.  The value of $\gamma$ obtained from unitarity
triangle fitting is in the range of $60^0 \sim 80^0$. But in
analysis of non-leptonic $B$ decay, possibility of larger $\gamma$
has been discussed by Deshpande {\it et al.} \cite{18} and He {\it
et al.} \cite{19}. The obtained value of $\gamma$ is $\gamma =90^0
\sim 140^0$.  In our calculations we use two representative values
of $\gamma = 110^0$ and $\gamma =65^0$.

In Tables III $-$ VI, we show the branching ratios and the CP
asymmetries for $B \rightarrow VT$ decays with either $\Delta S
=0$ or $|\Delta S| =1$.  In the tables the second and the third
columns correspond to the sets of the input parameters,
$$
\{\xi =0.1,~ m_s = 85~{\rm MeV},~ \gamma =110^0 \}~~~ {\rm and}~~~
\{\xi =0.1,~ m_s = 100~{\rm MeV},~ \gamma =65^0\}~,
$$
respectively. Similarly, the fourth
and the fifth columns correspond to the cases,
$$
\{\xi =0.3,~ m_s = 85~{\rm MeV},~ \gamma =110^0\}~~~{\rm and}~~~
\{\xi =0.3,~ m_s = 100~{\rm MeV},~ \gamma =65^0\}~,
$$
respectively.  The sixth and the seventh columns
correspond to the cases,
$$
\{\xi =0.5,~ m_s = 85~{\rm MeV},~ \gamma =110^0\}~~~{\rm and}~~~
\{\xi =0.5,~ m_s = 100~{\rm MeV}~, \gamma =65^0\}~,
$$
respectively.  Here $\xi \equiv 1/ N_c = 0.3$ corresponds to the
case of naive factorization ($N_c =3$).  It has been known that in
$B \rightarrow D$ decays the generalized factorization has been
successfully used with the favored value of $\xi \approx 0.5$
\cite{20}.  Also, as mentioned above, a recent analysis of
charmless $B$ decays to two light mesons such as $PP$ and $VP$
\cite{10} shows that $\xi \approx 0.45$ is favored with certain
values of other parameters for the best fit to the recent CLEO
data.

The branching ratios for $B \rightarrow VT$ decay modes with
$\Delta S =0$ are shown in Table III. Among $\Delta S=0$ modes,
the decay modes $B^+ \rightarrow \rho^+ a^0_2$, $B^+ \rightarrow
\rho^+ f_2$, and $B^0 \rightarrow \rho^+ a^-_2$ have relatively
large branching ratios of a few times $10^{-7}$.  The branching
ratio for $B^+ \to \rho^+ f_2^{\prime}$ is much smaller than that
for $B^+ \to \rho^+ f_2$ by about two orders of magnitude, because
the former decay rate is proportional to $\sin \phi_{_T} =0.13$,
instead of $\cos \phi_{T} =0.99$ which is a proportional factor of
the latter decay rate.  This prediction is consistent with that
based on flavor SU(3) symmetry. We see that in the factorization
scheme the following equality between the branching ratios holds
for any set of the parameters given above: $2 {\cal B}(B^+
\rightarrow \rho^+ a^0_2) \approx {\cal B}(B^0 \rightarrow \rho^+
a^-_2)$, as discussed in Eq. (\ref{facto1}). (Little deviation
from the exact equality arises from breaking of isospin symmetry.)
We also see from Table III that ${\cal B}(B^+ \rightarrow \rho^0
a^+_2)$ is much smaller than ${\cal B}(B^+ \rightarrow \rho^+
a^0_2)$ by an order of magnitude or even three orders of magnitude
depending on values of the input parameters.  This is because in
factorization the dominant contribution to the former mode arises
from the color-suppressed tree diagram ($C_T$) and further the
$C_T$ destructively interferes with $P_T$, while the dominant one
to the latter mode arises from the color-favored tree diagram
($T_T$) and the $T_T$ constructively interferes with $P_T$.  We
note that ${\cal B}(B^+ \rightarrow \rho^0 a^{+(0)}_2) \approx
{\cal B}(B^+ \rightarrow \omega a^{+(0)}_2)$ and ${\cal B}(B^+
\rightarrow \rho^0 f_2^{(\prime)}) \approx {\cal B}(B^+
\rightarrow \omega f_2^{(\prime)})$, as is expected from the fact
that $\rho^0$ and $\omega$ have the similar quark content and the
decay amplitudes for the modes having $\rho^0$ in the final state
are similar to those for the modes having $\omega$ in the final
state (some differences appear only in the penguin diagram
contributions which are small in $\Delta S =0$ decays).  The
branching ratios of most processes are order of $10^{-8}$ or less.
The CP asymmetries ${\cal A_{CP}}$ in $\Delta S=0$ decays are
shown in Table IV.  The CP asymmetries for $B^+ \rightarrow \rho^0
a^+_2$ and $B^+ \rightarrow \omega a^+_2$ can be as large as $27
\%$ and $49 \%$, respectively, with the branching ratio of
$O(10^{-8})$ for $\xi =0.5$~.

In Table VII, we show the ratio ${\cal B}(B \rightarrow VT) /
{\cal B} (B \rightarrow PT)$ for $\Delta S =0$ decays, where quark
contents of $V$ and $P$ are identical.  For comparison, we choose
the modes $B^+ \to \rho^+ a_2^0$ ~($B^+ \to \pi^+ a_2^0$), $B^+
\to \rho^+ f_2$ ~($B^+ \to \pi^+ f_2$), and $B^0 \to \rho^+ a_2^-$
~($B^0 \to \pi^+ a_2^-$) in $B \to VT$ ($B \to PT$) whose decay
amplitudes have the dominant tree diagram contribution $T_T$.  For
these modes, the ratio ${\cal B}(B \rightarrow VT) / {\cal B} (B
\rightarrow PT)$ can be written as
\begin{eqnarray}
{{\cal B}(B \rightarrow VT) \over {\cal B} (B \rightarrow PT)}
\approx { m_{_V} f^2_{_V} [{\cal X} |\vec p_{_V}|^7 +{\cal Y}
|\vec p_{_V}|^5 +{\cal Z} |\vec p_{_V}|^3 ] \over 2 |\vec
p_{_P}|^5 m_B^2 f_{_P}^2 [F^{B \to T} (m_{_P}^2)]^2 }~.
\label{vtpt}
\end{eqnarray}
In the ratio, the dependence on $G_F$, the CKM matrix elements,
and the effective coefficients $a_i$ does not appear.  The ratio
depends only on the form factors for $B \to T$ calculated in the
ISGW model, in addition to masses of $P$, $V$ and $T$, and the
decay constants $f_{_P}$ and $f_{_V}$.  Thus, the ISGW model and
the factorization scheme can be tested by measuring the above
ratio for different modes, as shown in Table VII, in future
experiment. Table VII shows that the ratio for $\Delta S =0$
decays are indeed insensitive to different values of the input
parameters, such as $\xi$ and the weak phase $\gamma$, and are in
between 0.473 and 0.495~.

The branching ratios and CP asymmetries for $|\Delta S|=1$ decay
processes are shown in Table V and VI, respectively. In $|\Delta
S|=1$ decays, the relevant penguin diagrams give dominant
contribution to the decay rates. We see that the branching ratios
for $|\Delta S|=1$ decays are in range between $O(10^{-7})$ and
$O(10^{-10})$, similar to those for $\Delta S =0$ decays. The
processes $B^+ \to K^{* +} a_2^0$, $K^{* +} f_2$, $K^{* 0} a_2^+$,
and $B^0 \to K^{* +} a_2^-$, $K^{* 0} a_2^0$, $K^{* 0} f_2$ have
relatively large branching ratios of $O(10^{-7})-O(10^{-8})$,
since the amplitudes for these modes have the dominant penguin
contribution $P_T^{\prime}$.  We note that the branching ratios
for $B \to \omega K_2^*$ and $B \to \phi K_2^*$ vary strongly
depending on $\xi$. This is mainly because the amplitudes for
these modes have the singlet penguin contribution $S_T^{\prime}$
and the magnitude of $S_T^{\prime}$ strongly depends on the value
of $\xi$ in the factorization scheme.  Unlike $\Delta S=0$ decays
such as $B \to \omega a_2$ and $B \to \phi a_2$, in $|\Delta S|
=1$ decays such as $B \to \omega K_2^*$ and $B \to \phi K_2^*$ the
tree contribution is suppressed compared to the penguin
contribution. Further, in the mode $B \to \omega K_2^*$, the
amplitude $2S_T^{\prime}$ is the only strong penguin contribution
so that the branching ratio for this mode varies strongly
depending on $\xi$ (even though $S_T^{\prime}$ is expected to be
small due to the OZI suppression).  In $B \to \phi K_2^*$, the
amplitude $P_T^{\prime} +S_T^{\prime}$ is the relevant strong
penguin contribution, and in factorization $S_T^{\prime}$ can
become comparable (with the opposite sign) to $P_T^{\prime}$ for
certain values of $\xi$, say, $\xi=0$ so that the branching ratio
for this mode strongly depends on $\xi$. Table VI shows the CP
asymmetries ${\cal A_{CP}}$ in $|\Delta S|=1$ decays. ${\cal
A_{CP}}$'s in most modes are expected to be small. In $B^+ \to
K^{* +} a_2^0$, $B^+ \to K^{* +} f_2$, and $B^0 \to K^{* +}
a_2^-$, ${\cal A_{CP}}$ can be about $15 \% - 25 \%$ with the
branching ratios of $O(10^{-7})-O(10^{-8})$.

In Table VII, we show the ratio ${\cal B}(B \rightarrow VT) /
{\cal B} (B \rightarrow PT)$ for the modes $B^+ \to K^{* +} a_2^0
~(B^+ \to K^+ a_2^0)$, $B^+ \to K^{* +} f_2 ~(B^+ \to K^+ f_2)$,
and $B^0 \to K^{* +} a_2^- ~(B^0 \to K^+ a_2^-)$ in $B \to VT$ ($B
\to PT$) whose amplitudes have the dominant penguin contribution
$P_T^{\prime}$.  For these modes, the ratio ${\cal B}(B
\rightarrow VT) / {\cal B} (B \rightarrow PT)$ can be {\it
approximately} expressed as Eq. (\ref{vtpt}), but unlike the
$\Delta S=0$ case, in this case, dependence of the ratio on the
weak phase $\gamma$ and the strange quark mass $m_s$ remains, due
to the effect of the suppressed tree diagram $T_T^{\prime}$ and
the $m_s$-dependence of ${\cal B} (B \rightarrow PT)$. In the
table, the second and the third columns correspond to the cases of
sets of the parameters: \{$m_s = 85$ MeV, $\gamma =110^0$\} and
\{$m_s = 100$ MeV, $\gamma =65^0$\}, respectively. In both cases,
the values of $\xi$ vary from 0.1 to 0.5~.  The result shows two
different ranges of values of the ratio: in the former case (the
second column), the ratio is about 2.5, while in the latter case
(the third column), the ratio is about 1.0~.  Given values of
$m_s$ and $\gamma$, the ratio is almost independent of the value
of $\xi$.

\section{Conclusion}

We have analyzed exclusive charmless decays $B \rightarrow VT$ in
the frameworks of both flavor SU(3) symmetry and generalized
factorization. Using the flavor SU(3) symmetry, we have shown that
certain decay modes, such as $B^+ \rightarrow \rho^+ a_2^0$,
$\rho^+ f_2$ and $B^0 \rightarrow \rho^+ a_2^-$ in $\Delta S =0$
decays, and $B^+ \to K^{* +} f_2$, $K^{* 0} a_2^+$ and $B^0 \to
K^{* +} a_2^-$ in $|\Delta S| =1$ decays, are expected to have the
largest decay rates and so these modes can be preferable to find
in future experiment. Certain ways to test validity of the
factorization scheme have been presented by emphasizing interplay
between both approaches and carefully combining the predictions
from both approaches.  We have also shown that $B$ meson decays to
two tensor mesons in the final state do not happen in the
factorization scheme, which can be tested in future experiment.

We have calculated the branching ratios and CP asymmetries for $B
\rightarrow VT$ decays, using the \emph{full} effective
Hamiltonian including all the penguin operators which are
essential to analyze the $|\Delta S| =1$ processes and to
calculate CP asymmetries.  We have also used the non-relativistic
quark model proposed by Isgur, Scora, Grinstein, and Wise to
obtain the form factors describing $B \rightarrow T$ transitions.
As shown in Tables III and V, the branching ratios vary from
$O(10^{-7})$ to $O(10^{-10})$. Consistent with the prediction from
the flavor SU(3) analysis, the decay modes such as $B^+
\rightarrow \rho^+ a_2^0$, $\rho^+ f_2$, $B^0 \rightarrow \rho^+
a_2^-$ and $B^{+(0)} \rightarrow K^{*0(+)}_2 a_2^{+(-)}$ have the
branching ratios of order of $10^{-7}$.  We have identified the
decay modes where the CP asymmetries are expected to be large,
such as $B^+ \rightarrow \rho^0 a^+_2$ and $B^+ \rightarrow \omega
a^+_2$ in $\Delta S =0$ decays, and $B^+ \to K^{* +} a_2^0$, $B^+
\to K^{* +} f_2$, and $B^0 \to K^{* +} a_2^-$ in $|\Delta S| =1$
decays. Due to possible uncertainties in the hadronic form factors
of $B \to VT$ and non-factorization effects, the predicted
branching ratios could be increased. We have also presented the
ratio ${\cal B}(B \rightarrow VT) / {\cal B} (B \rightarrow PT)$
for $\Delta S=0$ and $|\Delta S|=1$ decays, which primarily
depends on the form factors for $B \to T$, especially in $\Delta
S=0$ case.  Thus, measurement of this ratio for different modes in
future experiment can test the ISGW modes and the factorization
ansatz.  Although experimentally challenging, the exclusive
charmless decays, $B \to VT$, can probably be carried out in
details at hadronic $B$ experiments such as BTeV and LHC-B, where
more than $10^{12}$ $B$-mesons will be produced per year, as well
as at present asymmetric $B$ factories of Belle and Babar.
\\

\centerline{\bf ACKNOWLEDGEMENTS}
\medskip

\noindent The work of C.S.K., D.Y.K., H.S.K, and J.S.H was
supported by the BSRI Program of MOE, Project No. 99-015-D10032.
The work of B.H.L and S.O. was supported by the KRF Grants,
Project No. 2000-015-DP0077.


\newpage
\begin{center}
{\bf APPENDIX}
\end{center}

In this Appendix, we present expressions for all the decay
amplitudes of $B \rightarrow VT$ modes shown in Tables I and II as
calculated in the factorization scheme.  Below we use $F_{\alpha
\beta}^{B \rightarrow T}$ defined in Eqs. (\ref{FBT}).
\\
\noindent
(1) $B \rightarrow VT$ ($\Delta S = 0$) decays.
\begin{eqnarray}
A(B^+ \rightarrow \rho^+ a_2^{0}) &=& {G_F \over 2} ( m_{\rho^+}
f_{\rho^+} \epsilon^{* \alpha \beta} F_{\alpha \beta}^{B
\rightarrow {a_2^{0}}} (m^2_{\rho^+})) \left\{ V^*_{ub} V_{ud} a_1
- V^*_{tb} V_{td} (a_4 + a_{10}) \right\}
\\
A(B^+ \rightarrow \rho^+ f_2) &=&  {G_F \over 2} ( m_{\rho^+}
f_{\rho^+} \epsilon^{* \alpha \beta}  F_{\alpha \beta}^{B
\rightarrow {f_2}} (m^2_{\rho^+}))        \left\{ V^*_{ub}
V_{ud}ca_1  - V^*_{tb} V_{td} c(a_4 + a_{10}) \right\}
\\
A(B^+ \rightarrow  \rho^+ f_2^{\prime} ) &=&  {G_F \over 2} (
m_{\rho^+}  f_{\rho^+} \epsilon^{* \alpha \beta} F_{\alpha
\beta}^{B \rightarrow {f_2^{\prime}}} (m^2_{\rho^+})) \left\{
V^*_{ub} V_{ud} sa_1  - V^*_{tb} V_{td} s(a_4 + a_{10}) \right\}
\\
A(B^+ \rightarrow \rho^0 a_2^+) &=&  {G_F \over 2} ( m_{\rho^0}
f_{\rho^0} \epsilon^{* \alpha \beta}  F_{\alpha \beta}^{B
\rightarrow {a_2^+}} (m^2_{\rho^0}))   \left\{ V^*_{ub} V_{ud} a_2
\right.   \nonumber \\
&\mbox{}& \left. - V^*_{tb} V_{td} [-a_4 +{3 \over 2} (a_7+a_9)
+{1 \over 2}a_{10} ] \right\}
\\
A(B^+ \rightarrow \omega a_2^+) &=&  {G_F \over 2} ( m_{\omega}
f_{\omega} \epsilon^{* \alpha \beta}  F_{\alpha \beta}^{B
\rightarrow {a_2^+}} (m^2_{\omega}))   \left\{ V^*_{ub} V_{ud} a_2
\right.   \nonumber\\
&\mbox{}& \left. - V^*_{tb} V_{td} [2 (a_3+a_5) + a_4 +{1 \over 2}
(a_7 +a_9) -{1 \over 2}a_{10} ] \right\}
\\
A(B^+ \rightarrow \phi a_2^+) &=&  {G_F \over \sqrt{2}}   ( m_{
\phi}  f_{ \phi} \epsilon^{* \alpha \beta}  F_{\alpha \beta}^{B
\rightarrow {a_2^+}} (m^2_{ \phi}))        \left\{ - V^*_{tb}
V_{td} [(a_3+a_5) -{1 \over 2} (a_7 +a_9) ] \right\}
\\
A(B^+ \rightarrow \bar K^{* 0} K_2^{*+}) &=& {G_F \over \sqrt{2}}
( m_{ \bar K^{* 0}}  f_{ \bar K^{* 0}} \epsilon^{* \alpha \beta}
F_{\alpha \beta}^{B\rightarrow {K_2^{*+}}} (m^2_{ \bar K^{* 0}}))
\left\{  - V^*_{tb} V_{td} [a_4 -{1 \over 2}a_{10} ] \right\}
\\
A(B^+ \rightarrow \bar K^{* +} \bar K_2^{0}) &=& 0 \\
A(B^0 \rightarrow \rho^+ a_2^-) &=&  {G_F \over \sqrt{2}} (
m_{\rho^+}  f_{\rho^+} \epsilon^{* \alpha \beta}  F_{\alpha
\beta}^{B \rightarrow {a_2^-}} (m^2_{\rho^+}))   [V^*_{ub} V_{ud}
a_1  - V^*_{tb} V_{td} ( a_4+ a_{10}) ]
\\
A(B^0 \rightarrow \rho^-  a_2^+) &=& 0  \\
A(B^0 \rightarrow \rho^0 a_2^{0}) &=&  {G_F \over 2 \sqrt{2}} (
m_{\rho^0}  f_{\rho^0} \epsilon^{* \alpha \beta}  F_{\alpha
\beta}^{B \rightarrow {a_2^{0}}} (m^2_{\rho^0}))   \left\{
V^*_{ub} V_{ud} a_2
\right.   \nonumber\\
&\mbox{}& \left. - V^*_{tb} V_{td} [ -a_4 +{3 \over 2} (a_7+a_9)
+{1 \over 2}a_{10} ] \right\}
\\
A(B^0 \rightarrow \rho^0 f_2) &=&  {G_F \over 2 \sqrt{2}} (
m_{\rho^0}  f_{\rho^0} \epsilon^{* \alpha \beta}  F_{\alpha
\beta}^{B \rightarrow {f_2}} (m^2_{\rho^0}))   \left\{ V^*_{ub}
V_{ud} ca_2
\right.   \nonumber\\
&\mbox{}& \left. - V^*_{tb} V_{td} c[-a_4 +{3 \over 2} (a_7+a_9)
+{1 \over 2}a_{10}  ] \right\}           \\
A(B^0 \rightarrow \rho^0 f_2^{\prime}) &=&  {G_F \over 2 \sqrt{2}}
( m_{\rho^0 }  f_{\rho^0 } \epsilon^{* \alpha \beta} F_{\alpha
\beta}^{B \rightarrow {f_2^{\prime}}} (m^2_{\rho^0 })) \left\{
V^*_{ub} V_{ud}s a_2
\right.   \nonumber\\
&\mbox{}& \left. -V^*_{tb} V_{td} s[-a_4 +{3 \over 2} (a_7+a_9)
+{1 \over 2}a_{10}  ] \right\}                                \\
A(B^0 \rightarrow \omega a_2^{0}) &=&  {G_F \over 2 \sqrt{2}} (
m_{\omega}  f_{\omega} \epsilon^{* \alpha \beta}  F_{\alpha
\beta}^{B \rightarrow {a_2^{0}}} (m^2_{\omega}))   \left\{
V^*_{ub} V_{ud} a_2
\right.   \nonumber\\
&\mbox{}& \left. - V^*_{tb} V_{td} [2(a_3 +a_5) + a_4 +{1 \over 2}
(a_7+a_9) -{1 \over 2}a_{10}     ] \right\}
\\
A(B^0 \rightarrow \omega f_2) &=&  {G_F \over 2 \sqrt{2}} (
m_{\omega}  f_{\omega} \epsilon^{* \alpha \beta}  F_{\alpha
\beta}^{B \rightarrow {f_2}} (m^2_{\omega^0}))   \left\{ V^*_{ub}
V_{ud}c a_2
\right.   \nonumber\\
&\mbox{}& \left. - V^*_{tb} V_{td}c [2(a_3 +a_5) + a_4 +{1 \over
2} (a_7+a_9) -{1 \over 2}a_{10}     ] \right\}
\\
A(B^0 \rightarrow \omega f_2^{\prime}) &=&  {G_F \over 2 \sqrt{2}}
( m_{\omega }  f_{\omega } \epsilon^{* \alpha \beta} F_{\alpha
\beta}^{B \rightarrow {f_2^{\prime}}} (m^2_{\omega })) \left\{
V^*_{ub} V_{ud}s a_2
\right.   \nonumber\\
&\mbox{}& \left. - V^*_{tb} V_{td}s [2(a_3 +a_5) + a_4 +{1 \over
2} (a_7+a_9) -{1 \over 2}a_{10}  ]
\right\}                                     \\
A(B^0 \rightarrow \phi a_2^{0}) &=&  {G_F \over 2 } ( m_{\phi}
f_{\phi} \epsilon^{* \alpha \beta}  F_{\alpha \beta}^{B
\rightarrow {a_2^{0}}} (m^2_{\phi}))   \left\{ - V^*_{tb} V_{td}
[(a_3 +a_5) -{1 \over 2} (a_7+a_9)] \right\}
\\
A(B^0 \rightarrow \phi f_2) &=&  {G_F \over 2 } ( m_{\phi}
f_{\phi} \epsilon^{* \alpha \beta}  F_{\alpha \beta}^{B
\rightarrow {f_2}} (m^2_{\phi}))   \left\{ - V^*_{tb} V_{td}
c[(a_3 +a_5) -{1 \over 2} (a_7+a_9)] \right\}
\\
A(B^0 \rightarrow \phi f_2^{\prime}) &=&  {G_F \over 2 } (
m_{\phi}  f_{\phi} \epsilon^{* \alpha \beta}  F_{\alpha \beta}^{B
\rightarrow {f_2^{\prime}}} (m^2_{\phi}))   \left\{ -V^*_{tb}
V_{td} s[(a_3 +a_5) -{1 \over 2} (a_7+a_9)] \right\}
\\
A(B^0 \rightarrow \bar K^{* 0} K_2^{* 0}) &=& {G_F \over \sqrt{2}}
( m_{\bar K^{* 0}}  f_{\bar K^{* 0}} \epsilon^{* \alpha \beta}
F_{\alpha \beta}^{B \rightarrow {K_2^{* 0}}} (m^2_{\bar K^{* 0}}))
\left\{- V^*_{tb} V_{td} [ a_4  -{1 \over 2}a_{10} ] \right\}
\\
A(B^0 \rightarrow  K^{* 0}  \bar K_2^{* 0}) &=& 0
\end{eqnarray}  \\
\noindent (2) $B \rightarrow VT$ ($|\Delta S| = 1$) decays.
\begin{eqnarray}
A(B^+ \rightarrow K^{* +} a_2^{0}) &=&  {G_F \over  2} ( m_{K^{*
+}}  f_{K^{* +}} \epsilon^{* \alpha \beta}  F_{\alpha \beta}^{B
\rightarrow {a_2^{0}}} (m^2_{K^{* +}})) \left\{V^*_{ub} V_{us} a_1
-V^*_{tb} V_{ts} (a_4 +a_{10}) \right\} \\
A(B^+ \rightarrow K^{* +} f_2 ) &=&  {G_F \over  2} ( m_{K^{* +}}
f_{K^{* +}} \epsilon^{* \alpha \beta}  F_{\alpha \beta}^{B
\rightarrow {f_2}} (m^2_{K^{* +}}))   \left\{V^*_{ub} V_{us} ca_1
-V^*_{tb} V_{ts}c (a_4 +a_{10} ) \right\} \\
A(B^+ \rightarrow K^{* +} f_2^{\prime}) &=&{G_F \over  2} (
m_{K^{* +}}  f_{K^{* +}} \epsilon^{* \alpha \beta}  F_{\alpha
\beta}^{B \rightarrow T} (m^2_{V}))   \left\{V^*_{ub} V_{us} sa_1
-V^*_{tb} V_{ts}s (a_4 +a_{10} ) \right\} \\
A(B^+ \rightarrow K^{* 0} a_2^{ +}) &=&  {G_F \over \sqrt{2}} (
m_{K^{* 0} }  f_{K^{* 0} } \epsilon^{* \alpha \beta} F_{\alpha
\beta}^{B \rightarrow {a_2^{ +}}} (m^2_{K^{* 0} })) \left\{
-V^*_{tb} V_{ts} (a_4 -{1 \over 2}a_{10} ) \right\}    \\
A(B^+ \rightarrow \rho^+ K_2^{* 0}) &=& 0 \\
A(B^+ \rightarrow \rho^0 K_2^{* +}) &=&  {G_F \over 2 } (
m_{\rho^0}  f_{\rho^0} \epsilon^{* \alpha \beta}  F_{\alpha
\beta}^{B \rightarrow {K_2^{* +}}} (m^2_{\rho^0})) \left\{V^*_{ub}
V_{us} a_2  - V^*_{tb} V_{ts} {3 \over 2} (a_7+a_9) \right\}
\\
A(B^+ \rightarrow \omega K_2^{* +}) &=&  {G_F \over 2} (
m_{\omega}  f_{\omega} \epsilon^{* \alpha \beta}  F_{\alpha
\beta}^{B \rightarrow {K_2^{* +}}} (m^2_{\omega})) \left\{V^*_{ub}
V_{us} a_2  - V^*_{tb} V_{ts} [2(a_3 +a_5) +{1 \over 2} (a_7+a_9)
] \right\}
\\
A(B^+ \rightarrow \phi K_2^{* +}) &=&  {G_F \over \sqrt{2}} (
m_{\phi}  f_{\phi} \epsilon^{* \alpha \beta}  F_{\alpha \beta}^{B
\rightarrow {K_2^{* +}}} (m^2_{\phi}))   \left\{ -V^*_{tb} V_{ts}
[a_3 +a_4 +a_5  -{1 \over 2} (a_7+a_9+a_{10}) ] \right\}
\\
A(B^0 \rightarrow K^{* +} a_2^{-}) &=&  {G_F \over \sqrt{2} } (
m_{K^{* +}}  f_{K^{* +}} \epsilon^{* \alpha \beta}  F_{\alpha
\beta}^{B \rightarrow {a_2^{-}}} (m^2_{K^{* +}}))   \left\{
V^*_{ub} V_{us}
a_1 -{V_{tb}}^* V_{ts} (a_4 +a_{10}) \right\} \\
A(B^0 \rightarrow K^{* 0} a_2^{0}) &=&  {G_F \over 2 } ( m_{K^{*
0}}  f_{K^{* 0}} \epsilon^{* \alpha \beta}  F_{\alpha \beta}^{B
\rightarrow {a_2^{0}}} (m^2_{K^{* 0}}))   \left\{- V^*_{tb} V_{ts}
(a_4 -{1 \over 2}a_{10}) \right\} \\
A(B^0 \rightarrow K^{* 0} f_2 ) &=&  {G_F \over 2 } ( m_{K^{* 0}}
f_{K^{* 0}} \epsilon^{* \alpha \beta}  F_{\alpha \beta}^{B
\rightarrow {f_2}} (m^2_{K^{* 0}}))   \left\{  -V^*_{tb} V_{ts}
c(a_4 -{1 \over 2}a_{10}) \right\} \\
A(B^0 \rightarrow K^{* 0} f_2^{\prime}) &=& {G_F \over 2 } (
m_{K^{* 0}}  f_{K^{* 0}} \epsilon^{* \alpha \beta} F_{\alpha
\beta}^{B \rightarrow {f_2^{\prime}}} (m^2_{K^{* 0}})) \left\{
-{V^*_{tb}}  V_{ts}  s(a_4 -{1 \over 2}a_{10}) \right\} \\
A(B^0 \rightarrow \rho^{-} K_2^{* +}) &=&  0
\\
A(B^0 \rightarrow \rho^0 K_2^{* 0 }) &=&  {G_F \over 2 } (
m_{\rho^0}  f_{\rho^0} \epsilon^{* \alpha \beta}  F_{\alpha
\beta}^{B \rightarrow {K_2^{* 0 }}} (m^2_{\rho^0}))
\left\{V^*_{ub} V_{us} a_2  - V^*_{tb} V_{ts} {3 \over 2}
(a_9+a_7) \right\}
\\
A(B^0 \rightarrow \omega K_2^{* 0 }) &=&  {G_F \over 2 } (
m_{\omega}  f_{\omega} \epsilon^{* \alpha \beta}  F_{\alpha
\beta}^{B \rightarrow {K_2^{* 0 }}} (m^2_{\omega}))
\left\{V^*_{ub} V_{us} a_2 -V^*_{tb} V_{ts} [ 2(a_3 +a_5) +{1
\over 2} (a_7+a_9) ] \right\}
\\
A(B^0 \rightarrow \phi K_2^{* 0 }) &=&  {G_F \over \sqrt{2}} (
m_{\phi }  f_{\phi } \epsilon^{* \alpha \beta}  F_{\alpha
\beta}^{B \rightarrow {K_2^{* 0 }}} (m^2_{\phi }))   \left\{
-V^*_{tb} V_{ts} [a_3 +a_4 +a_5  -{1 \over 2} (a_7+a_9+a_{10}) ]
\right\}
\end{eqnarray}

\newpage
\begin{table}
\caption{Coefficients of SU(3) amplitudes in $B \rightarrow VT$
($\Delta S = 0$). The coefficients of the SU(3) amplitudes with
the subscript $V$ are expressed in square brackets.  As explained
in Sec. II, the contributions of the SU(3) amplitudes with the
subscript $V$ vanish in the framework of factorization, because
those contributions contain the matrix element $\langle T|
J^{\rm{weak}}_{\mu} |0 \rangle$, which is zero.  Here $c$ and $s$
denote $\cos \phi_{_T}$ and $\sin \phi_{_T}$, respectively. }
\begin{tabular}{c|ccccccc}
$B \rightarrow VT$ & factor & $T_T$$[T_V]$ & $C_T$$[C_V]$ &
$S_T$$[S_V]$ & $P_T$$[P_V]$ & $P_{EW,T}$ $[P_{EW,V}]$&
$P_{EW,T}^C$$[P_{EW,V}^C]$
\\ \hline
$B^+ \rightarrow \rho^+ a_2^0$ & $-{1 \over \sqrt{2}}$ & 1 & [1] &
0 & $1,~[-1]$ & [1] & ${2 \over 3}$, $\left[ {1 \over 3} \right]$
\\ $B^+ \rightarrow \rho^+ f_2$ & ${1 \over \sqrt{2}}$ & $c$ & $[c]$ &  $[2c+ \sqrt{2} s]$ &
$c,~[c]$ & $\left[{c-\sqrt{2}s}\over 3 \right]$ &${2c \over 3}$,
$\left[ -{c  \over 3} \right]$
\\ $B^+ \rightarrow \rho^+ f_2^{\prime}$ & $1 \over \sqrt{2}$ & $s$ & $[s]$ & $[2s- \sqrt{2} c]$ &
$s,~[s]$ & $\left[{\sqrt{2} c +s}\over 3 \right]$ &${2s \over 3}$,
$\left[ -{s  \over 3} \right]$
\\ $B^+ \rightarrow \rho^0 a_2^+$ & $-{1 \over \sqrt{2}}$ & [1] & 1 &
0 & $-1,~[1]$ & 1 & $1 \over 3$, $\left[ 2 \over3 \right] $
\\ $B^+ \rightarrow \omega a_2^+$ & $1 \over \sqrt{2}$ & $[1]$ & $1$ & $2$ & $1,~[1]$
&${1 \over 3} $ & $-{1 \over 3}$, $\left[ 2 \over 3 \right]$
\\ $B^+ \rightarrow \phi a_2^+$ & 1 & 0 & 0 & 1 & 0 &$ -{1 \over 3}$ & $0$
\\ $B^+ \rightarrow K^{* +} \bar K_2^{* 0}$ & 1 & 0 & 0 & 0 & [1] & 0 & $[-{1 \over 3}]$
\\ $B^+ \rightarrow \bar K^{* 0} K_2^{* +}$ & 1 & 0 & 0 & 0  & 1 & 0 & $-{1 \over 3}$

\\ $ B^0 \rightarrow \rho^+ a_2^ -$ & $-1$ & 1 & 0 & 0 & 1 & 0 & $2 \over 3$
\\ $B^0 \rightarrow \rho^- a_2^+ $  & $-1$ &  [1] & 0 & 0 & [1] & 0 & $\left[2 \over 3\right]$
\\ $ B^0 \rightarrow \rho^0 a_2^0$ & $-{1 \over 2}$ & 0 & $1$, $[1]$ &0 & $-1$, $[-1]$ &$1,~[1]$
& ${1 \over 3}$, $\left[ {1 \over 3} \right]$
\\ $  B^0 \rightarrow \rho^0 f_2$  &$-{1 \over 2}$ & 0 & $c,~[ -c]$ & $[-(2c+ \sqrt{2} s)]$
& $-c$, $[-c]$ & $c,~\left[  {-c + \sqrt{2} s}\over 3  \right]$ &
${c \over 3}$, $\left[{c \over3} \right]$
\\ $  B^0 \rightarrow \rho^0 f_2^{\prime}$ &$-{1 \over 2}$ & 0 & $s,~[-s]$
& $[-(2s- \sqrt{2} c)]$ & $-s$, $[-s]$ & $s,~\left[-{ (\sqrt{2} c
+s)}\over 3\right] $ & $s \over 3$, $\left[{s \over3} \right]$
\\ $ B^0 \rightarrow \omega a_2^0$ & ${1 \over 2}$ & 0 & $1,~[-1]$  & $2$
& $1,~ [1]$ & ${1 \over 3}$, $[-1]$ & $-{1 \over 3},~ \left[ -{1
\over 3} \right]$
\\ $  B^0 \rightarrow \omega f_2$  &$1 \over 2$ & 0 & $c,~[c]$ & $2c$, $[(2c+ \sqrt{2} s)]$
& $c,~[c]$ & ${c \over 3},~ \left[{c- \sqrt{2} s}\over 3\right]$ &
$-{c \over 3}$,  $[-{c \over3}]$
\\ $ B^0 \rightarrow \omega f_2^{\prime}$ &$1 \over 2$ & 0 & $s,~[s]$ & $2s,~ [(2s- \sqrt{2} c)]$
& $s,~ [s]$ & ${s \over 3},~ \left[{{s+\sqrt{2} c}\over 3}\right]$
& $-{s \over 3},~ [-{s \over3}]$
\\ $ B^0 \rightarrow \phi a_2^0$ & ${1 \over \sqrt{2}}$ & 0 & 0& $1$ & 0 & $-{1 \over 3}$ &0
\\ $ B^0 \rightarrow \phi f_2$ & $1 \over \sqrt{2}$ & 0 & 0& c & 0 & $-{c \over 3}$ &0
\\ $ B^0 \rightarrow \phi f_2^{\prime}$ & $1 \over \sqrt{2}$ & 0 & 0& s & 0 & $-{s \over 3}$ &0
\\ $  B^0 \rightarrow  K^{* 0} \bar K_2^{* 0}$ & 1 & 0 & 0 & 0 & [1] & 0 & $\left[-{1 \over 3}\right]$
\\ $  B^0 \rightarrow \bar K^{* 0} K_2^{* 0}$ & 1 & 0 & 0 & 0 & 1 & 0 & $-{1 \over 3}$
\end{tabular}
\end{table}


\newpage
\begin{table}
\caption{Coefficients of SU(3) amplitudes in $B \rightarrow VT$ (
$|\Delta S| = 1$ ).}
\begin{tabular}{c|ccccccc}
$B \rightarrow VT$ & factor & $T_T^{\prime}$$[T_P^{\prime}]$ &
$C_T^{\prime}$$[C_V^{\prime}]$ & $S_T^{\prime}$$[S_V^{\prime}]$ &
$P_T^{\prime}[P_V^{\prime}]$ & $P_{EW,T}^{\prime}$
$[P_{EW,V}^{\prime}]$& $P_{EW,T}^{C \prime}$$[P_{EW,V}^{C
\prime}]$
\\ \hline
$B^+ \rightarrow K^{* +} a_2^0$ &$-{1 \over \sqrt{2}}$  & 1 & [1]
& 0 & 1 & [1] & $2 \over 3$
\\ $B^+ \rightarrow K^{* +} f_2$ & ${1 \over \sqrt{2}}$ & $c$ & $[c]$ &
$[2c+ \sqrt{2} s]$ & $c,~[ \sqrt{2} s]$ & $\left[{c- \sqrt{2}
s}\over 3 \right]$ & ${2 \over 3}c$, $\left[-{\sqrt{2}s \over
3}\right]$
\\ $B^+ \rightarrow K^{* +} f_2^{\prime}$ & ${1 \over \sqrt{2}}$ & $s$ & $[s]$
& $[2s- \sqrt{2} c]$ & $s,~[- \sqrt{2} c]$ & $\left[{s+ \sqrt{2}
c}\over 3 \right]$ & ${2 \over 3}s$, $\left[{\sqrt{2}c \over
3}\right]$
\\ $B^+ \rightarrow   K^{* 0} a_2^+$ & 1 & 0 & 0 & 0 & 1 & 0 & $-{1 \over 3}$
\\ $B^+ \rightarrow \rho^+  K_2^{* 0}$ & 1 & 0 & 0 & 0 & [1] & 0 &
$\left[-{1 \over 3} \right]$
\\ $B^+ \rightarrow \rho^0 K_2^{* +}$    &$-{1 \over \sqrt{2}}$  & $[1]$
& 1 & 0 & [1] & 1 & $\left[2 \over 3\right]$
\\ $B^+ \rightarrow \omega K_2^{* +}$   &$1 \over \sqrt{2}$  & $[1]$ & 1 & 2
& [1] &$ 1 \over 3$ & $\left[2 \over 3 \right]$
\\ $B^+ \rightarrow \phi K_2^{* +}$    &1 & $0$ & 0 & 1 & 1 &
$-{1 \over 3}$ & $-{1 \over 3}$

\\ $ B^0 \rightarrow K^{* +} a_2^-$  & $-1$ & 1&0 & 0 & 1 &0 &$2 \over 3$
\\ $ B^0 \rightarrow  K^{* 0} a_2^0$ & ${1 \over \sqrt{2}}$ & 0& $[-1]$ & 0 & $1$
& $[-1]$ &$-{1\over 3}$
\\ $ B^0 \rightarrow  K^{* 0} f_2$ & $1 \over \sqrt{2}$ & 0 & $[c]$
& $[2c+\sqrt{2} s]$ & $c,~[\sqrt{2} s] $ & $\left[{c-\sqrt{2}
s}\over 3\right]$&$- {c\over 3} $, $[- {\sqrt{2} s\over 3 }]$
\\ $ B^0 \rightarrow  K^{* 0} f_2^{\prime}$ & $1 \over \sqrt{2}$ & 0 & $[s]$
& $[2s-\sqrt{2} c]$ & $s,~[-\sqrt{2} c]$ & $\left[{s+\sqrt{2}
c}\over 3 \right]$ &$- {s\over 3} $, $[ {\sqrt{2} c\over 3 }]$
\\$ B^0 \rightarrow \rho^- K_2^{* +}$ & $-1$ & [1] & $0$ & 0 & [1]
&0& $\left[ {2\over3} \right]$
\\ $ B^0 \rightarrow \rho^0  K_2^{* 0}$ &$-{1 \over \sqrt{2}}$  & 0 & 1 & 0
& $[-1]$  & 1& $\left[ 1 \over 3\right]$
\\ $ B^0 \rightarrow \omega  K_2^{* 0}$ & $1 \over \sqrt{2}$ & 0 & 1 & 2
& [1] & $1 \over 3$& $\left[-{1 \over 3}\right]$
\\ $ B^0 \rightarrow \phi  K_2^{* 0}$ & 1 & 0 & 0 & 1 & 1 & $- {1\over 3}$
&$- {1\over 3}$
\end{tabular}
\end{table}

\begin{table}
\caption{The branching ratios for $B \rightarrow VT$ decay modes
with $\Delta S =0$. The second and the third columns correspond to
the cases of sets of the parameters: \{$\xi =0.1$, $m_s = 85$ MeV,
$\gamma =110^0$\} and \{$\xi =0.1$, $m_s = 100$ MeV, $\gamma
=65^0$\}, respectively. Similarly, the fourth and the fifth
columns corresponds to the cases: \{$\xi =0.3$, $m_s = 85$ MeV,
$\gamma =110^0$\} and \{$\xi =0.3$, $m_s = 100$ MeV, $\gamma
=65^0$\}, respectively.  The sixth and the seventh columns
correspond to the cases: \{$\xi =0.5$, $m_s = 85$ MeV, $\gamma
=110^0$\} and \{$\xi =0.5$, $m_s = 100$ MeV, $\gamma =65^0$\},
respectively.}
\smallskip
\begin{tabular}{c|cccccc}
Decay mode                                      & ${\cal
B}(10^{-8})$& ${\cal B}(10^{-8})$&${\cal B}(10^{-8})$ &${\cal
B}(10^{-8})$ &  ${\cal B}(10^{-8})$ & ${\cal B}(10^{-8})$
\\ \hline
  $B^+ \rightarrow \rho^+ a_2^0$ &21.93 &22.17 &19.46 &19.70
&17.13 &17.37
\\  $B^+ \rightarrow \rho^+ f_2$ &23.33 &23.58 &20.70 &20.95 &18.23
&18.48
\\ $B^+ \rightarrow \rho^+ f_2^{\prime}$ &0.26 &0.26 &0.23 &0.23 &0.20 &0.20
\\  $B^+ \rightarrow \rho^0 a_2^+$ &0.84 &0.78
&0.046 &0.033 &1.10 &1.16
\\ $B^+ \rightarrow \omega a_2^+$ &0.77 &0.77 &0.039 &0.034 &1.18
&1.28
\\  $B^+ \rightarrow \phi a_2^+$ &0.064 &0.053 &0.006 &0.006 &0.022 &0.012
\\  $B^+ \rightarrow \bar K^{* 0} K_2^{* +}$ &0.062 &0.041 &0.053 &0.033
&0.045 &0.027
\\  $ B^0 \rightarrow \rho^+ a_2^ -$ &40.72 &41.16 &36.13 &36.57 &31.81
&32.26
\\ $ B^0 \rightarrow \rho^0 a_2^0$ &0.39 &0.36 &0.022 &0.015 &0.51
&0.54
\\ $  B^0 \rightarrow \rho^0 f_2$ &0.42 &0.38 &0.023 &0.016 &0.55 &0.57
\\ $  B^0 \rightarrow \rho^0 f_2^{\prime}$ &0.005 &0.004 &0.0003 &0.0002 &0.006
&0.006
\\ $ B^0 \rightarrow \omega a_2^0$ &0.36 &0.36 &0.018 &0.016 &0.55
&0.60
\\ $  B^0 \rightarrow \omega f_2$ &0.38 &0.38 &0.019 &0.017 &0.58
&0.63
\\ $ B^0 \rightarrow \omega f_2^{\prime}$ &0.004 &0.004 &0.0002 &0.0002
&0.006 &0.007
\\ $ B^0 \rightarrow \phi a_2^0$ &0.030 &0.025 &0.003 &0.003 &0.010
&0.006
\\ $ B^0 \rightarrow \phi f_2$ &0.030 &0.025 &0.003 &0.003 &0.010
&0.006
\\ $ B^0 \rightarrow \phi f_2^{\prime}$ &0.0004 &0.0003 &0 &0 &0.0001 &0
\\ $  B^0 \rightarrow \bar K^{* 0} K_2^{* 0}$ &0.12 &0.076 &0.098 &0.062 &0.082
&0.050
\end{tabular}
\end{table}


\begin{table}
\caption{The CP asymmetries for $B \rightarrow VT$ decay modes
with $\Delta S =0$.  The definitions for the columns are the same
as those in Table III. }
\smallskip
\begin{tabular}{c|cccccc}
Decay mode                  &${\cal A_{CP}}$ &${\cal A_{CP}}$
&${\cal A_{CP}}$& ${\cal A_{CP}}$ & ${\cal A_{CP}}$& ${\cal
A_{CP}}$
\\ \hline
  $B^+ \rightarrow \rho^+ a_2^0$ &$-0.073$ &$-0.070$ &$-0.072$
&$-0.069$ &$-0.071$ &$-0.068$
\\  $B^+ \rightarrow \rho^+ f_2$ &$-0.073$ &$-0.070$ &$-0.072$ &$-0.069$
&$-0.071$ &$-0.068$
\\ $B^+ \rightarrow \rho^+ f_2^{\prime}$ &$-0.073$ &$-0.070$ &$-0.072$
&$-0.069$ &$-0.071$ &$-0.068$
\\  $B^+ \rightarrow \rho^0 a_2^+$
&$-0.34$ &$-0.36$ &0.66 &0.91 &0.27 &0.25
\\ $B^+ \rightarrow \omega a_2^+$ & 0.017 &0.016 &$-0.72$ &$-0.79$ &$-0.49$
&$-0.44$
\\  $B^+ \rightarrow \phi a_2^+$  &0 &0 &0 &0 &0 &0
\\  $B^+ \rightarrow \bar K^{* 0} K_2^{* +}$  &0 &0 &0 &0 &0 &0
\\  $ B^0 \rightarrow \rho^+ a_2^ -$ &$-0.073$ &$-0.070$ &$-0.072$ &$-0.069$
&$-0.071$ &$-0.068$
\\ $ B^0 \rightarrow \rho^0 a_2^0$  &$-0.34$ &$-0.36$ &0.66 &0.91 &0.27 &0.25
\\ $  B^0 \rightarrow \rho^0 f_2$ &$-0.34$ &$-0.36$ &0.66 &0.91 &0.27 &0.25
\\ $  B^0 \rightarrow \rho^0 f_2^{\prime}$ &$-0.34$ &$-0.36$ &0.66 &0.91 &0.27 &0.25
\\ $ B^0 \rightarrow \omega a_2^0$ &0.017 &0.016 &$-0.72$ &$-0.79$ &$-0.49$
&$-0.44$
\\ $  B^0 \rightarrow \omega f_2$  &0.017 &0.016 &$-0.72$ &$-0.79$ &$-0.49$
&$-0.44$
\\ $ B^0 \rightarrow \omega f_2^{\prime}$ &0.017 &0.016 &$-0.72$ &$-0.79$ &$-0.49$
&$-0.44$
\\ $ B^0 \rightarrow \phi a_2^0$  &0 &0 &0 &0 &0 &0
\\ $ B^0 \rightarrow \phi f_2$  &0 &0 &0 &0 &0 &0
\\ $ B^0 \rightarrow \phi f_2^{\prime}$  &0 &0 &0 &0 &0 &0
\\ $  B^0 \rightarrow \bar K^{* 0} K_2^{* 0}$   &0 &0 &0 &0 &0 &0
\end{tabular}
\end{table}

\newpage
\begin{table}
\caption{The branching ratios  for $B \rightarrow VT$ decay modes
with $|\Delta S| =1$.  The definitions for the columns are the
same as those in Table III.}
\smallskip
\begin{tabular}{c|cccccc}
Decay mode                                     & ${\cal
B}(10^{-8})$& ${\cal B}(10^{-8})$&${\cal B}(10^{-8})$ &${\cal
B}(10^{-8})$& ${\cal B}(10^{-8})$ & ${\cal B}(10^{-8})$
\\ \hline
  $B^+ \rightarrow K^{* +} a_2^0$ &10.78 &5.97 &9.74 &5.40
&8.75 &4.88
\\  $B^+ \rightarrow K^{* +} f_2$ &11.20 &6.19 &10.11 &5.61 &9.09
&5.06
\\  $B^+ \rightarrow K^{* +} f_2^{\prime}$ &0.14 &0.078 &0.13 &0.070 &0.11
&0.064
\\  $B^+ \rightarrow   K^{* 0} a_2^+$ &16.45 &16.45 &12.97 &12.97 &9.91 &9.91
\\   $B^+ \rightarrow \rho^0 K_2^{* +}$ &0.59 &0.81
&0.57 &0.55 &0.62 &0.39
\\  $B^+ \rightarrow \omega K_2^{* +}$ &5.30 &4.70 &0.029 &0.035
&3.91 &3.28
\\  $B^+ \rightarrow \phi K_2^{* +}$ &2.52 &2.52 &10.39 &10.39 &23.66
&23.66
\\ $ B^0 \rightarrow K^{* +} a_2^-$ &20.48 &11.33 &18.50 &10.27 &16.62
&9.26
\\ $ B^0 \rightarrow  K^{* 0} a_2^0$ &7.65 &7.65 &6.03 &6.03 &4.61 &4.61
\\ $ B^0 \rightarrow  K^{* 0} f_2$ &7.94 &7.94 &6.26 &6.26 &4.78 &4.78
\\  $ B^0 \rightarrow  K^{* 0} f_2^{\prime}$ &0.10 &0.10 &0.079 &0.079 &0.060 &0.060
\\ $ B^0 \rightarrow \rho^0  K_2^{* 0}$  &0.54  &0.75 &0.53 &0.50 &0.57 &0.36
\\  $ B^0 \rightarrow \omega  K_2^{* 0}$  &4.87 &4.32 &0.027 &0.032
&3.60 &3.02
\\   $ B^0 \rightarrow \phi  K_2^{* 0}$ &2.34 &2.34 &9.64 &9.64 &21.96
&21.96
\end{tabular}
\end{table}

\begin{table}
\caption{The CP asymmetries for $B \rightarrow VT$ decay modes
with $|\Delta S| =1$.  The definitions for the columns are the
same as those in Table III.}
\smallskip
\begin{tabular}{c|cccccc}
Decay mode                                      &${\cal A_{CP}}$
&${\cal A_{CP}}$ &${\cal A_{CP}}$& ${\cal A_{CP}}$ & ${\cal
A_{CP}}$& ${\cal A_{CP}}$
\\ \hline
  $B^+ \rightarrow K^{* +} a_2^0$  &$-0.15$& $-0.26$& $-0.14$ &
$-0.25$ &$-0.14$  &$-0.24$
\\  $B^+ \rightarrow K^{* +} f_2$  &$-0.15$& $-0.26$& $-0.14$ & $-0.25$ &$-0.14$  &$-0.24$
\\ $B^+ \rightarrow K^{* +} f_2^{\prime}$  &$-0.15$& $-0.26$& $-0.14$ & $-0.25$ &$-0.14$  &$-0.24$
\\  $B^+ \rightarrow   K^{* 0} a_2^+$  &0 &0 &0 &0 &0 &0
\\   $B^+ \rightarrow \rho^0 K_2^{* +}$
&$-0.006$ &$-0.004$ &0.001 &0.001 &0.007 &0.010
\\  $B^+ \rightarrow \omega K_2^{* +}$  &$-0.035$ &$-0.038$ &0.107 &0.088 &$-0.041$  &$-0.047$
\\  $B^+ \rightarrow \phi K_2^{* +}$  &0 &0 &0 &0 &0 &0
\\ $ B^0 \rightarrow K^{* +} a_2^-$ &$-0.15$ &$-0.26$ &$-0.14$ &$-0.25$ &$-0.14$ &$-0.24$
\\ $ B^0 \rightarrow  K^{* 0} a_2^0$  &0 &0 &0 &0 &0 &0
\\ $ B^0 \rightarrow  K^{* 0} f_2$  &0 &0 &0 &0 &0 &0
\\  $ B^0 \rightarrow  K^{* 0} f_2^{\prime}$ &0 &0 &0 &0 &0 &0
\\ $ B^0 \rightarrow \rho^0  K_2^{* 0}$  &$-0.006$ &$-0.004$ &0.001 &0.001 &0.007 &0.010
\\  $ B^0 \rightarrow \omega  K_2^{* 0}$  &$-0.035$ &$-0.038$ &0.107 &0.088 &$-0.041$  &$-0.047$
\\   $ B^0 \rightarrow \phi  K_2^{* 0}$  &0 &0 &0 &0 &0 &0
\end{tabular}
\end{table}

\begin{table}
\caption{Ratios of the branching ratios for $B \to VT$ and for $B
\to PT$ decay modes, where $V$ and $P$ have identical quark
content. The second and the third columns correspond to the cases
of sets of the parameters: \{$m_s = 85$ MeV, $\gamma =110^0$\} and
\{$m_s = 100$ MeV, $\gamma =65^0$\}, respectively. In both cases,
the values of $\xi$ vary from 0.1 to 0.5~. }
\smallskip
\begin{tabular}{ccc}
Ratio & $m_s = 85$ MeV, $\gamma =110^0$ & $m_s = 100$ MeV, $\gamma
=65^0$
\\ \hline
${\cal B}(B^+ \rightarrow \rho^+ a_2^0$) / ${\cal B}(B^+
\rightarrow \pi^+ a_2^0$)  & 0.482$-$0.483 & 0.495
\\  ${\cal B}(B^+ \rightarrow \rho^+ f_2$) / ${\cal B}(B^+ \rightarrow
\pi^+ f_2$)  & 0.472$-$0.473 & 0.484$-$0.485
\\ ${\cal B}(B^0 \rightarrow \rho^+ a_2^-$) / ${\cal B}(B^0 \rightarrow
\pi^+ a_2^-$)  & 0.473$-$0.474 & 0.485$-$0.486
\\ ${\cal B}(B^+ \rightarrow K^{* +} a_2^0$) / ${\cal B}(B^+ \rightarrow
K^+ a_2^0$)  & 2.50$-$2.55 & 1.03$-$1.10
\\ ${\cal B}(B^+ \rightarrow K^{* +} f_2$) / ${\cal B}(B^+ \rightarrow
K^+ f_2$)  & 2.39$-$2.50 & 0.99$-$1.05
\\ ${\cal B}(B^0 \rightarrow K^{* +} a_2^-$) / ${\cal B}(B^0 \rightarrow
K^+ a_2^-$)  & 2.51$-$2.63 & 1.04$-$1.10
\end{tabular}
\end{table}
\end{document}